\magnification = \magstep1
\hsize = 16 true cm
\vsize = 22 true cm
\baselineskip = 24 true pt
\centerline {\bf {A CLASS OF QUANTUM INTEGRABLE MODELS ASSOCIATED WITH}} 
\centerline {\bf {THE INFRA-RED LIMIT OF MASSIVE CHERN-SIMONS THEORY }} 
\medskip 
\centerline {${\bf V.~ V.~ Sreedhar}^*$}
\centerline {S. N. Bose National Centre for Basic Sciences}
\centerline {JD Block, Sector 3, Bidhannagar}
\centerline {Calcutta 700091, India}
\bigskip 
\centerline {${\bf Abstract}$}

We study the infra-red limit of non-abelian Chern-Simons gauge theory 
perturbed by a non-topological, albeit gauge invariant, mass term. It 
is shown that, in this limit, we may construct an infinite class of 
integrable quantum mechanical models which, for the case of SU(2) 
group, are labelled by the angular momentum eigenvalue. The first 
non-trivial example in this class is obtained for the triplet 
representation and it physically describes the gauge invariant  
coupling of a non-abelian Chern-Simons particle with a particle  
moving on $S^3$ - the SU(2) group manifold. In addition to this, the model  
has a fascinating resemblance to the Landau problem and may be regarded 
as a non-abelian and  a non-linear generalisation of the same 
defined on the three-sphere with the uniform magnetic field replaced  
by an angular momentum field. We explicitly solve for some eigenstates   
of this model in a closed form in terms of some generalised orthogonal  
polynomials. In the process, we unravel some startling connections  
with Anderson's chain models which are important in the study of  
disordered systems in condensed matter physics. We also sketch a 
method which allows us, in principle, to find the energy eigenvalues 
corresponding to the above eigenstates of the theory if the Lyapunov 
exponents of the transfer matrix of the infinite chain model involved  
are known.  
\vfill
\hrule
\smallskip 
* E-mail address: sreedhar@bose.ernet.in\hfil\break  
\vfil\eject 
Gauge theories with Chern-Simons terms [1] have spanned a wide range of 
interests in the past decade or so. The impressive array of topics making 
use of ideas related to Chern-Simons gauge theories extends, on the one hand, 
from the purely mathematical - pertaining to issues of the topology of three 
manifolds [2] - to the phenomenological, on the other, as in models of 
quantum Hall effect [3] which are testable in the laboratory. From a more 
formal point of view, Chern-Simons gauge theories have shed light on aspects 
of anyonic spin and statistics, and conformally invariant quantum field 
theories in two space dimensions [4]. Driven by a desire to understand some 
aspects of such theories in the simpler setting of 0+1 dimensions, models of 
abelian Chern-Simons quantum mechanics have also been constructed, canonically 
quantised, and solved [5, 6]. These models are described by Lagrangians 
which are quantum mechanical analogues of the so-called self-dual models  
in 2+1 dimensions with and without a Maxwell term [7]. The equation of   
motion of one of these models is given by the famous Lorentz equation  
for a charged particle moving in an external electromagnetic field. As is  
well known, the spectrum of this model is described by Landau levels [8]. 
The other model, obtained by tuning a certain dimensionful coupling 
to zero, corresponds to projecting on to the lowest Landau level.  
As such, both the models are of immediate relevance in quantum Hall effect. 

The conventional way of thinking of the Chern-Simons term is to regard it as a  
gauge invariant mass term in the lagrangian density for a gauge theory in 2+1 
dimensions with the kinetic piece being given, as usual, by the Maxwell or 
the Yang-Mills term [1]. Contrary to this, one may also construct theories in 
which the Chern-Simons term plays the role of the kinetic term while a Proca  
term takes the place of a mass term [7]. This is the abelian self-dual model  
mentioned above. As described in [6], the infra-red limit of such a theory  
yields one of the Chern-Simons quantum mechanics models discussed in [5].   
In the present paper we consider a non-abelian extension of this model and 
show that we may construct an infinite class of quantum integrable models 
which, for the case of SU(2) group, are labelled by the eigenvalue of the
angular momentum operator. The first non-trivial model in this class will
be discussed in some detail and explicit expressions will be worked out  
for some of its eigenstates. We will also sketch a method which allows us to
calculate, in principle, the corresponding energy eigenvalues. 

To begin with, let us recall some basic notions and establish notations. The 
non-abelian Chern-Simons gauge theory is defined by the lagrangian density 
$${\cal L}_{CS} = {k\over 4\pi}\epsilon^{\mu\nu\lambda} Tr (A_\mu\partial_\nu
A_\lambda + i{2\over 3} A_\mu A_\nu A_\lambda )\eqno(1)$$
where $A_\mu = A_\mu^aT_a,~ T_a$ being traceless, hermitian matrices in the 
fundamental representation of the gauge group SU(N). The $T_a$ satisfy the 
following algebra:
$$ [T_a, T_b]_- = if_{ab}^cT_c,~ [T_a, T_b]_+ = d_{ab}^cT_c + 
{1\over N}\delta_{ab},~ 
Tr (T_aT_b) = {\delta_{ab}\over 2} $$
The metric $h_{\mu\nu} = diag (-1, 1, 1)$, the completely antisymmetric 
Levi-Civita tensor is such that  
$\epsilon^{0ij} = \epsilon^{ij} = -\epsilon_{ij}$ with 
$\epsilon^{012} = 1$ and we sum over repeated indices without comment. 
>From simple dimensional considerations it follows that $A_\mu$ has dimensions 
of mass and, consequently, the coupling $k$ is dimensionless. 
Under a gauge transformation, 
$$ A_\mu (x)\rightarrow A_\mu^U(x) = U^{-1}(x)A_\mu (x)U(x) - 
iU^{-1}(x)\partial_\mu U(x)\eqno(2)$$
where $U(x) \in SU(N)$, the action transforms as follows:
$$S[A] \rightarrow \tilde S[A^U] = S[A] + 2\pi kn \eqno(3a)$$
where 
$$ n = {1\over 24\pi^2} \int d^3x~ \epsilon^{\mu\nu\lambda} Tr 
(U^{-1}\partial_\mu U U^{-1}\partial_\nu U U^{-1}\partial_\lambda U)\eqno(3b)$$ 
is the winding number of the map 
$$U(x) : M_3 \rightarrow SU(N)\eqno(3c)$$
$M_3$ being the compact three-manifold on which the theory is defined. Thus, 
although the action is not gauge invariant, the generating function would be 
so if the dimensionless Chern-Simons coupling $k$ is quantised to be an 
integer. 

Obviously, a mass term $A_\mu A^\mu $ for the gauge field is gauge 
non-invariant. Let us, however, consider the following term  
$$ {\cal L}_m = -m Tr [(A_\mu + i V^{-1}\partial_\mu V)(A^\mu + i V^{-1}
\partial^\mu V)] \eqno(4)$$
where the SU(N)-valued auxiliary field $V$ is designed to transform in such 
a way that the whole term is gauge invariant. $m$ has dimensions of mass and 
the auxiliary field $V$ is dimensionless. The total action  
$S = \int d^3x~ {\cal L} = \int d^3x~ ({\cal L}_{CS} + {\cal L}_m )$ 
is then gauge invariant 
under the simultaneous gauge transformations 
$$A_\mu \rightarrow A_\mu^U = U^{-1}A_\mu U -iU^{-1}\partial_\mu U\eqno(5a)$$
and 
$$V\rightarrow V^U = VU \eqno(5b)$$
In $S$ only terms linear and quadratic in $A_0$ appear. We may 
therefore readily do the $A_0$ integration in the path integral after 
doing a Wick rotation $t\rightarrow it$. After the $A_0$ integration,  
the lagrangian density takes the form
$$\eqalignno {{\cal L} &= 
-{m\over 2}[(A_i^a + (iV^{-1}\partial_iV)^a)
(A_{ia} + (iV^{-1}\partial_iV)_a) + 
(V^{-1}\partial_0V)^a(V^{-1}\partial_0V)_a]\cr 
&- {k^2\over 32\pi^2m}[F_{ij}^a + {4\pi m\over k}(V^{-1}\partial_0V)^a]
[F_{ija} + {4\pi m\over k}(V^{-1}\partial_0V)_a] 
 - {k\over 8\pi }\epsilon^{ij}A_i^a\delta_{ab}\partial_0 A_j^b }$$
The $F_{ij}$ in the above equation is the usual field strength tensor 
defined by 
$$F^a_{ij} = \partial_iA^a_j - \partial_jA^a_i - f^a_{bc}A^b_iA^c_j $$ 
In order to be able to pick the relevant degrees of freedom of each of 
the fields appearing in the above terms in the infra-red limit, it is   
useful to put the system in a box. The dimensions of the box $L$ will
subsequently be taken to infinity. Introduce, therefore,
$$A_i ({\bf x}, t) = {1\over L}\sum_{\bf n} \exp {2\pi i {\bf x}\cdot {\bf n}
\over L}
q_{i{\bf n}} (t) \eqno(6a) $$
and 
$$V({\bf x}, t) =\sum_{\bf n}
\exp {2\pi i{\bf x}\cdot {\bf n}\over L}v_{\bf n} (t)\eqno(6b)$$
In the long wavelength, or the infra-red, limit all the massive modes will be  
suppressed and we can approximate the sums on the right hand sides of the  
above equations by just the zero momentum components. In what follows, the  
spatially constant modes thus obtained for $A_i$ and $V$ are simply denoted 
by $q_i$ and $v$ respectively. If we now take the limit  
$L\rightarrow \infty $ in the action and drop all the irrelevant terms
we get 
$$S = \int dt~[-{k\over 8\pi}\epsilon^{ij} q_i^a {\cal D}_{ab} q_j^b - 
{m\over 2}  
q_i^aq_{ia} - m (v^{-1}\partial_0v)^a(v^{-1}\partial_0v)_a]\eqno(7a)$$
where ${\cal D}$ is the covariant derivative in the adjoint representation
$${\cal D}_{ab} = \delta_{ab}\partial_0 - f_{ab}^c(v^{-1}\partial_0v)_c 
\eqno(7b)$$
Introducing local coordinates $\xi^a$ on the group manifold we now have 
$$ Tr (v^{-1}\partial_0v)^2 = g_{ab}(\xi)\dot\xi^a\dot\xi^b $$
where $g_{ab}(\xi)$ is the metric induced on the manifold by the map in 
Eq.(3c), and is given by
$$g_{ab}(\xi) =  Tr (v^{-1}{\partial v\over \partial \xi^a}
v^{-1}{\partial v\over \partial \xi^b})\eqno(8)$$ 
The action now takes the form 
$$S = \int dt~[ -{k\over 8\pi}\epsilon^{ij} q_i^a{\cal D}_{ab}q_j^b -  
{m\over 2}q_i^aq_{ia} - 2m g_{ab}(\xi)\dot\xi^a\dot\xi^b ]\eqno(9a)$$
The covariant derivative in the above equation is written in terms of 
$\xi^a$ as
$${\cal D}_{ab}=\delta_{ab}\partial_0-f_{ab}^cg_{cd}(\xi)\dot\xi^d\eqno(9b)$$ 
Let us first consider the $\xi$-independent part of the Lagrangian which 
we denote by $L_0$.
$$ L_0 = -{k\over 8\pi} \epsilon^{ij} q_i^a \dot q_{ja} - {m\over 2} 
q_i^aq_{ia} \eqno(10)$$
The momentum conjugate to $q$ is then given by 
$$p_j^a \equiv {\partial L\over\partial\dot q_{ja}} =
 {k\over 8\pi}\epsilon_{ij}q_i^a\eqno(11a)$$
The above equation yields the following second class constraints 
$$\chi_j^a\equiv p_j^a - {k\over 8\pi}\epsilon_{ij}q_i^a\approx  0\eqno(11b)$$ 
It is straightforward to evaluate the Dirac bracket [9] for such a 
constrained system and it yields 
$$\{q_i^a, q_j^b\}_{DB} = {4\pi\over k}\epsilon_{ji}\delta^{ab}\eqno(12a)$$
We can now work out the canonical momentum as  
$$ p_i^a = {k\over 4\pi}\epsilon_{ji}q_j^a \eqno(12b)$$
An identical expression is arrived at directly by using the Faddeev-Jackiw 
formalism [9].
The Hamiltonian is given by the usual Legendre prescription as follows
$$H_0 = p_i^a\dot q_i^a - L = m ({4\pi\over k})^2
\left( {p_1^ap_1^a\over 2} + ({k\over 4\pi})^2 {q_1^aq_1^a\over 2}
\right)\eqno(13)$$
if we eliminate one of the degrees of freedom using the definition of the  
canonical momentum in Eq.(12b) to arrive at the second equality. Thus, the 
$\xi$-independent part of the Hamiltonian just represents a bunch of $(N^2-1)$ 
harmonic oscillators. This is very much like in abelian Chern-Simons 
quantum mechanics of [5,6]. It is curious, though, to note in passing, 
that the non-abelian feature of the problem is reflected in the fact that 
the frequency of each of these oscillators is quantised in units of 
${1\over 4\pi}$. 

We now turn our attention to the more interesting $\xi$-dependent part of
the Lagrangian which we denote by $L_\xi$. 
$$L_\xi = -2m g_{ab}(\xi )\dot\xi^a\dot\xi^b + {k\over 8\pi } 
\epsilon^{ij}q_i^aq_j^bf_{ab}^cg_{cd}(\xi )\dot\xi^d \eqno(14)$$
Once again, the momentum canonically conjugate to $\xi$ can be written as  
$$\pi^a \equiv {\partial L\over\partial\dot\xi_a} = -4m\dot\xi^a + 
{k\over 8\pi}\epsilon^{ij}q_i^bq_j^cf^a_{bc}\eqno(15)$$  
The corresponding Hamiltonian works out to be
$$H_\xi = -{1\over 2m} g_{ab} (\xi ){\cal P}^a{\cal P}^b \eqno(16a)$$
where    
$${\cal P}^a = {1\over 2}(\pi^a - {k\over 8\pi}\epsilon^{ij}q_i^cq_j^d
f^a_{cd})\eqno(16b)$$
Physically, this part of the Hamiltonian just represents the dynamics of a  
particle moving on the SU(N) group manifold. The extra term in ${\cal P}^a$ is, 
however, reminiscent of the gauge field in the magnetic translation operator 
of the Landau problem [8]. Indeed, in this case it leads to a gauge invariant 
coupling with the non-abelian Chern-Simons particle whose Hamiltonian is 
given in Eq.(13). It may be appropriate to mention at this stage that $H_\xi$
commutes with $H_0$. Since the latter is trivially solved, we shall henceforth 
concentrate our efforts exclusively on the $\xi$-dependent part of the theory.  

To get a better insight into the problem, it is prudent, at this 
juncture, for us to specialise to the case of SU(2). An element    
$v \in SU(2)$ can always be represented in terms of Euler angles 
$\alpha ,\beta ,\gamma $ and the Pauli matrices as follows:
$$v = \exp (i\sigma_3 {\alpha\over 2})\exp (i\sigma_2 {\beta\over 2})
\exp (i\sigma_3 {\gamma\over 2})\eqno(17a)$$
where 
$$0\leq\alpha < 2\pi ,~ 0\leq\beta\leq\pi ,~ -2\pi <\gamma < 2\pi\eqno(17b)$$
Now let 
$$\xi^1\equiv\alpha ,~ \xi^2 \equiv\beta ,~ \xi^3 \equiv\gamma \eqno(18)$$
The metric on the SU(2) group manifold can then be calculated using the 
formula in Eq.(8) and works out to be 
$$ g_{ab} = -{1\over 2}\pmatrix {1&0&\cos\beta\cr
0&1&0\cr \cos\beta &0&1\cr}\eqno(19)$$ 
The determinant of the metric is easily shown to be 
$$det~ \mid g_{ab}\mid = -{1\over 2} \sin^2\beta\eqno(20)$$
The metric is therefore singular at $\beta = 0$ and $\beta = \pi$. These 
two points, for which the analysis needs to be done separately, will be 
ignored in the following. 
Substituting the metric in the expression for $H_\xi$ we get 
$$ H_\xi = {1\over 4m}\{\eta_{ab}{\cal P}^a{\cal P}^b + \cos\beta 
({\cal P}^1{\cal P}^3 + {\cal P}^3{\cal P}^1)\} \eqno(21)$$
where $\eta_{ab} $ is the flat metric. 
Using the fact that for $SU(2)$, $f^{abc} = \epsilon^{abc}$ we may recast 
the expression for ${\cal P}$ as 
$$ {\cal P}^a = {1\over 2} (\pi^a - 2L^a) \eqno(22)$$
where $L^a$ are the components of the angular momentum operator. It is useful 
to record the commutation relations satisfied by the operators in $H_\xi$
$$[\xi^a, \xi^b] = [\pi^a, \pi^b] = 0,~[\xi^a, \pi^b] =
i\hbar g^{ab}(\xi)\eqno(23a)$$
and 
$$[{\cal P}^a, {\cal P}^b] = i\hbar\epsilon^{abc}L^c = {i\hbar\over 2}
\epsilon^{abc}
(\pi^c - 2P^c) \eqno(23b)$$
It is interesting to note that the system has three degrees of freedom 
{\it viz.}, $\alpha , \beta$, and $ \gamma $. However, $\alpha$ and $\gamma$  
are cyclic coordinates. Hence, it also has three constants of motion,  
namely, $\pi^\alpha , \pi^\gamma$, and the Hamiltonian, $H_\xi$.  
It is therefore integrable. In fact, since ${\cal P}^a$ depends 
on the angular momentum operator, $L^a$, we have one integrable system 
for each representation of the Casimir ${\bf L}^2$. This is the infinite 
class of quantum integrable models advertised in the title of the paper 
and alluded to earlier.  

Since $\pi^\alpha$ and $\pi^\gamma$ are constants of motion, let us, 
without further ado, set them equal to zero. The Hamiltonian, henceforth 
to be simply denoted by $H$, reduces to  
$$H = {1\over 4m}{\bf L}^2 - {1\over 4m}\pi L_z + {1\over 16m} 
\pi^2 + {1\over 4m} [L_y, L_x]_+ \cos\beta \eqno(24)$$
where we have identified $L^1, L^2,$ and $L^3$ with the components of the 
angular momentum $L_y, L_z,$ and $L_x$ respectively and dropped the suffix 
on the second component of the momentum for elegance. With ${\bf L}^2$ 
and $L_z$ being diagonal, as usual, we get for the Hamiltonian in the  
singlet representation, $H_s$,  
$$H_s = {1\over 16m}\pi^2 \eqno(25)$$
{\it i.e.}, it is just the Hamiltonian for a free particle. The triplet 
representation presents the simplest non-trivial model in the 
infinite class of integrable models presented above and will now be discussed. 
Introduce then the angular momentum matrices for the triplet representation of 
SU(2) group. 
$$ \eqalign {L_x &= {1\over \sqrt2}\pmatrix {0&1&0\cr 1&0&1\cr 0&1&0\cr},~
 L_y = {1\over \sqrt2}\pmatrix {0&-i&0\cr i&0&-i\cr 0&i&0\cr},\cr 
 L_z &= \pmatrix {1&0&0\cr 0&0&0\cr 0&0&-1\cr},~ 
{\bf L}^2 = 2\pmatrix {1&0&0\cr 0&1&0\cr 0&0&1\cr}\cr}$$
Plugging in these matrices in the equation for $H$ we get for the Hamiltonian
in the triplet representation, $H_t$,   
$$H_t = \pmatrix {{\pi^2\over 16m} - {\pi\over 4m}&0&
-{i\over 4m}\cos\beta\cr 0&{\pi^2\over 16m}&0\cr
{i\over 4m}\cos\beta &0&{\pi^2\over 16m} +
 {\pi\over 4m}\cr} \eqno(26)$$
The Schr\"odinger equation can then be written as 
$$\pmatrix {-{\hbar^2\over 16m}d_\beta^2 + {i\hbar^2\over 4m}
d_\beta &
0&-{i\over 4m}\cos\beta\cr 0&-{\hbar^2\over 16m}d_\beta^2&0\cr
+{i\over 4m }\cos\beta &0& -{\hbar^2\over 16m}d_\beta^2 - 
{i\hbar^2\over 4m}d_\beta\cr} \pmatrix {\psi_1(\beta )\cr
\psi_2(\beta )\cr\psi_3(\beta )} = \epsilon \pmatrix {\psi_1(\beta )\cr
\psi_2(\beta )\cr\psi_3(\beta )}\eqno(27) $$
The equation for $\psi_2$ 
$$-{\hbar^2\over 16m}d_\beta^2 \psi_2 (\beta )
 = \epsilon \psi_2(\beta )\eqno(28a) $$
is easily solved to get 
$$ \psi_2 = C_1\exp (i{4\beta\over \hbar}\sqrt {m \epsilon})
+ C_2\exp (-i{4\beta\over \hbar}\sqrt {m \epsilon})\eqno(28b) $$
$C_1$ and $C_2$ being arbitrary constants. 
This is very much like in the Landau problem [8] where the component of the 
wavefunction along the direction of the magnetic field is just given by 
a plane wave solution. The non-trivial physics resides in the plane 
perpendicular to the magnetic field. It is also well-known that, in this 
plane, the problem can be mapped into the problem of a particle in a 
harmonic oscillator potential. As already mentioned, the present problem  
is a non-linear generalisation of the above situation in the sense that the
harmonic oscillator gives way to the mathematical pendulum. Further,
because of the non-abelian nature of the problem, we have a matrix mathematical
pendulum to contend with. With this understanding we may now return to the  
case at hand {\it i.e.,} obtaining the other two components of $\psi (\beta )$
by solving the two simultaneous differential equations 
$$(d_\beta^2 - 4id_\beta )\psi_1 (\beta ) + 4i
\cos\beta\psi_3 (\beta ) = E\psi_1 (\beta ) \eqno(29a)$$
$$ (d_\beta^2 + 4id_\beta )\psi_3 (\beta ) - 4i 
\cos\beta\psi_1 (\beta ) = E\psi_3 (\beta )\eqno(29b)$$
where 
$$E = -{16m\over\hbar^2}\epsilon\eqno(30)$$
This is a set of second order linear differential equations with periodic 
coefficients. Hence the solution we are looking for can be expanded in a 
Fourier series as 
$$\psi_1(\beta ) = \sum_{n=-\infty}^\infty a_n\exp (in\beta ) \eqno(31a)$$
$$\psi_3(\beta ) = \sum_{n=-\infty}^\infty b_n\exp (in\beta ) \eqno(31b)$$
where the coefficients $a_n$ and $b_n$ are, in general, complex.   
It is worth mentioning here that the solutions of a system of differential 
equations with periodic coefficients need not, in general, be periodic [10]. 
Indeed, for a generic period, the wavefunction is only required to return to  
itself modulo a phase which is determined by an appropriate unitary 
representation of the translation group. By setting this phase equal to the 
identity we are effectively restricting ourselves to the centre of 
each energy band in the Bloch picture. It is only for this subset of 
eigenstates that we will be able to obtain some exact results.  
Substituting the above expansions in Eq.(29) we get, after collecting 
terms, and setting the coefficient of each Fourier mode to zero, 
$$ -f(n)a_n + 2i(b_{n-1} + b_{n+1})
 = 0\eqno(32a)$$
and 
$$ -f(-n)b_n - 2i(a_{n-1} + a_{n+1})
 = 0 \eqno(32b)$$
where 
$$ f(n) = n^2 - 4n +E \eqno(32c)$$
The above two equations can be recast as a single matrix equation in terms 
of a two component column vector $A_n$. 
$$A_n = \pmatrix {a_n\cr b_n}\eqno(33)$$
After some algebra, this equation can be written in the following compact 
form 
$$A_{n-1} + A_{n+1} + {\bf M}_nA_n = 0 \eqno(34)$$
the matrix ${\bf M}_n$ being given by the equation 
$$ {\bf M}_n = {f(n) + f(-n)\over 4}\sigma_2 + 
{f(n) - f(-n)\over 4}\sigma_2\sigma_3 \eqno(35)$$
Before we embark on solving this equation, it is worth recalling what 
we are interested in. In terms of $A_n$, the two components of the 
wavefunction we are trying to solve for can be assembled into a 
column vector, which, for convenience, we shall denote by $\psi $.
We therefore have 
$$ \psi (\beta ) = \sum_{n=-\infty}^\infty A_n \exp (in\beta )\eqno(36a) $$
and  
$$ \psi^\dagger (\beta )=\sum_{n=-\infty}^\infty 
A_n^\dagger\exp (-in\beta )\eqno(36b) $$
Notice now that the system of Eq.(29) is invariant under the discrete 
transformations 
$$\beta \rightarrow -\beta ,~ \psi_1 \rightarrow \psi_3 ,~
\psi_3 \rightarrow -\psi_1 \eqno(37)$$
Hence, if $\psi_1(\beta )$ and $\psi_3(\beta )$ is a set of solutions of 
the above equations, so is the set $\psi_3(-\beta )$ and $-\psi_1(-\beta )$. 
Once again, as in the discussion following Eq. (31), the latter set could 
differ from the former by a constant multiplicative matrix. If we restrict 
ourselves to solutions which are strictly symmetric under the above discrete
transformations, however, we have,   
$$\pmatrix {\psi_1 (-\beta )\cr \psi_3(-\beta )\cr} =
-i\sigma_2 \pmatrix {\psi_1(\beta )\cr \psi_3 (\beta )\cr}\eqno(38a) $$
This can be 
further translated into a relation between $A_n$ and $A_{-n}$ as follows. 
$$A_{-n} = -i\sigma_2 A_n \eqno(38b)$$ 
Substituting the above results in Eqs.(36) we get 
$$ \psi (\beta ) = \sum_{n=0}^\infty [({\bf 1}  -i\sigma_2 )A_n \cos n\beta 
+ i({\bf 1} + i\sigma_2 )A_n \sin n\beta ]$$
and 
$$ \psi^\dagger (\beta ) = \sum_{n=0}^\infty [A_n^\dagger 
({\bf 1} - i\sigma_2 )^\dagger\cos n\beta -iA_n^\dagger
({\bf 1} + i\sigma_2 )^\dagger\sin n\beta ] $$
It easily follows from the above two equations that 
$$\int_0^\pi \psi^\dagger (\beta )\psi (\beta )d\beta 
= 4\pi \sum_{n=0}^\infty A_n^\dagger A_n = 4\pi\sum_{n=0}^\infty
\left( \mid a_n\mid^2 +\mid b_n\mid^2\right) \eqno(39)$$
We shall argue that the above sum is, in general, convergent. Hence the  
wavefunction is square integrable. We shall also derive closed expressions
for the above set of eigenfunctions.  

In Eq. (32a, b) let us eliminate one variable, say $b_n$. The resulting 
equation for $a_n$ couples $a_n$ with $a_{n-2}$ and $a_{n+2}$. This is 
significant because the even and odd coefficients completely decouple and 
may be treated separately. Since we are trying to solve a three-term 
recursion relation we need to fix two coefficients. We may set all   
the odd coefficients to zero by choosing the first two of them to vanish.  
Let us then deal only with the even coefficients. It is 
straightforward to show that these coefficients obey the recursion relation 
$$ \mu_{p-1}\alpha_{p-1} + \mu_p\alpha_{p+1} + \rho_p\alpha_p = 0 \eqno(40a)$$ 
where 
$$\mu_p = {1\over f(-(2p+1))}\eqno(40b)$$
and 
$$\rho_p = {{f(-(2p+1)) + f(-(2p-1)) - f(2p)f(-(2p+1))f(-(2p-1))}\over 
4f(-(2p-1))f(-(2p+1))}\eqno(40c)$$
and 
$$\alpha_p \equiv a_{2p}\eqno(40d)$$
If, on the contrary, we choose to set all the even coefficients to zero, 
the odd coefficients
$$\beta_p \equiv a_{2p+1}$$
satisfy a similar equation with slightly modified parameters 
$\mu$ and $\rho$. Identical considerations apply for the coefficients 
$b_n$.  
 
We now introduce a one (complex) parameter family of the above three-term
recursion relation as shown below and appeal to two theorems \footnote{*}
{These theorems, due to E. Hellinger and R. Nevanlinna, are discrete 
analogues of Weyl's theorems concerning Sturm-Liouville differential  
equations on the semi-infinite real line.}  
in the theory of orthogonal 
polynomials. The proofs of these theorems, well-known in mathematics 
literature, can be found in [11]. 

\noindent ${\bf Theorem~1}$: For any non-real $\lambda $, there exists at least 
one solution $\{\alpha_n (\lambda )\}_0^\infty $ of the equation 
$$\mu_{n-1}\alpha_{n-1}(\lambda ) + \mu_n\alpha_{n+1} (\lambda )
+ \rho_n\alpha_n (\lambda ) = \lambda\alpha_n (\lambda ) $$ 
for which 
$$\sum_{n=0}^\infty \mid\alpha_n (\lambda )\mid^2 < \infty  $$

The above equation actually has two linearly independent solutions 
by virtue of the fact that it is a second order linear difference 
equation. These solutions $P_n (\lambda )$ and $Q_n (\lambda )$
which satisfy the conditions 
$$P_0 (\lambda ) = 1,~ P_1 (\lambda ) = {\lambda - \rho_0\over \mu_0 }$$
and 
$$Q_0 (\lambda ) = 0,~ Q_1 (\lambda ) = {1\over \mu_0 }$$
are called orthogonal polynomials of the first and second kind 
respectively. Hence, let $\alpha_n (\lambda )$ be given by $P_n (\lambda )$.
Then,\hfil\break  
${\bf Theorem~2}$: If the series $\sum_{n=0}^\infty\mid P_n (\lambda )\mid^2$ 
converges at any non-real point  
$\lambda $, then it converges uniformly in every finite part of the 
complex $\lambda $-plane. An identical result also holds for $Q_n (\lambda )$.

It then follows immediately, by choosing the finite part of the complex plane
to be the origin, and $\alpha_n$ to be one or the other of the two types of
orthogonal polynomials mentioned above, that  
$\sum_{n=0}^\infty \mid\alpha_n (0)\mid^2 < \infty $. The convergence 
of the series $\sum\mid\beta_n(0)\mid^2$ can be 
argued analogously. Note further that for $\lambda = 0 $, the one parameter 
family of recursion relations we have introduced collapses to the recursion 
relation we are trying to solve.  
Since identical considerations also apply for $b_n$s, 
we may conclude that the sums on the right hand side of Eq. (39) are 
in general convergent and hence, the wavefunction is square integrable.   

The $\alpha_n $ or $\beta_n $, hence $a_n$, and similarly $b_n$, and 
consequently $A_n$, 
can be obtained in a closed form. In order to do this let us consider a 
related problem. This problem is specified by an infinite sequence of 
orthonormal orbitals $\{u_0,~u_1,~u_2, \cdots\}$ and a set of real parameters  
$\{\rho_0 ,\rho_1 , \rho_2 , \cdots \mu_0 , \mu_1 , \mu_2 ,\cdots\}$ 
which describe the 
action of the Hamiltonian ${\cal H}$ on the orbitals by a symmetric 
three-term recursion relation [12, 13]
$${\cal H}u_n = \rho_nu_n + \mu_nu_{n+1} + \mu_{n-1}u_{n-1} \eqno(41)$$ 
The $\alpha_n $ will be related to the eigenstates of the above 
Hamiltonian in a very specific way. 
The chain model defined above has a nice physical interpretation. The
orbital $u_0$ represents the initial state of the system and it could 
be, for example, an electron on a particular atom in a solid. The 
chain model Hamiltonian allows not only a certain on-site probability 
for each of the orbitals that are available, but also an amplitude for  
hopping from one site on the one dimensional lattice to the nearest and the 
next nearest neighbour sites. The example we are trying to solve 
is a particularly complicated chain model in the sense that both the hopping 
amplitude and the on-site amplitude are inhomogeneous.   
Let us, however, press ahead with our programme, undeterred, 
for a while. We begin by introducing a basis for the 
orbitals $u_n$. This basis is specified by representing $u_n$ as an infinite  
component vector all of whose elements except the $n$th are zero. The non-zero  
element is, further, chosen to be one. Remembering that $n$ runs from $0$ to 
$\infty $, then 
$$u_0 = \pmatrix {1\cr 0\cr \cdot\cr\cdot\cr\cdot\cr } ~ 
u_1 = \pmatrix {0\cr 1\cr \cdot\cr\cdot\cr\cdot\cr } ~~ \cdots \eqno(42)$$
It follows rather easily that the Hamiltonian ${\cal H}$ is expressed in 
this basis as a Jacobi (or a tridiagonal) matrix  
$$ {\cal H} = \pmatrix {\rho_0&\mu_0&~&~&~&~\cr \mu_0&\rho_1&\mu_1&~
&~&~\cr 
~&\mu_1&\rho_2&\mu_2&~&~\cr ~&~&\cdot &\cdot &\cdot &~\cr
~&~&~&\cdot &\cdot &\cdot\cr
~&~&~&~&\cdot &\cdot }\eqno(43)$$ 
In this representation it is obvious why the chain Hamiltonian is 
called a symmetric chain Hamiltonian. An eigenstate of the Hamiltonian 
is some linear combination of the states $\{u_0,~ u_1,~\cdots \}$ 
denoted below by a deliberate misuse of notation as 
$$\phi = \sum_{n=0}^\infty \alpha_n u_n \eqno(44)$$
such that 
$${\cal H}\phi = \lambda\phi \eqno(45)$$       
The above time-independent Schr\"odinger equation can be written in 
a matrix form as 
$$\pmatrix {(\lambda - \rho_0 )&-\mu_0&~&~&~&~\cr
-\mu_0& (\lambda -\rho_1 )&-\mu_1&~&~&~\cr
~&-\mu_1& (\lambda - \rho_2 )&-\mu_2&~&~\cr
~&~&-\mu_2& (\lambda - \rho_3 )&~&~\cr 
~&~&~&\cdot\cdot\cdot\cr 
~&~&~&~&\cdot\cdot\cr 
~&~&~&~&~&\cdot\cr }\pmatrix {\alpha_0\cr\alpha_1\cr\alpha_2\cr\cdot\cr\cdot
\cr\cdot\cr } = 0 \eqno(46)$$ 
The eigenvalues of the Hamiltonian ${\cal H}$ are given by the zeroes of the 
determinant $\Delta (\lambda )$ of the above matrix. The Jacobi form of the 
above matrix gives rise to a very simple recursion relation for its determinant.
It is easy to show that if we define $\Delta_{-1} (\lambda ) = 0$ and 
$\Delta_0 (\lambda ) = 1 $ then,  
$$\mu_n\Delta_{n+1} (\lambda ) +\mu_{n-1}\Delta_{n-1} (\lambda ) + 
\rho_n\Delta_n (\lambda ) 
= \lambda\Delta_n (\lambda ) \eqno(47)$$ 
The determinants appearing in the above equation are polynomials in $\lambda $
with the same order as their subscripts. If we now substitute for $\phi$ in 
Eq.(45) and collect coefficients of each orthogonal orbital $u_n$, we get 
$$\mu_{n-1}\alpha_{n-1}^{(p)} + \mu_n\alpha_{n+1}^{(p)} + \rho_n\alpha_n^{(p)} =
\lambda_p \alpha_n^{(p)}\eqno(48)$$
where $\lambda_p$ is a zero of $\Delta_N(\lambda )$ and hence an eigenvalue of 
the $N$-chain, $N$ being a generic integer which will be finally taken to
$\infty$. $\alpha_n^{(p)}$ defines an 
eigenstate corresponding to an eigenvalue $\lambda_p$ through the 
relation 
$${\cal H}\phi^{(p)} = \lambda_p\phi^{(p)} $$
By comparing Eq.(47) and (48) we get, apart from a common normalization,  
$$\alpha_n^{(p)} = \Delta_n (\lambda_p)\eqno(49) $$
If we now assume that there always exists an eigenvalue $\lambda_p = 0$, 
$$\alpha_n \equiv \Delta_n (0) \eqno(50)$$ 
yield solutions of Eq.(40a) that we had initially embarked upon solving. 
A few words regarding the  
properties of the above determinants are now in order. First, it  
is obvious that the secular determinants of the type discussed above would be  
finite for any model with $N$ points on the chain where $N$ is a generic, but  
finite integer. Since we are finally interested in solving an infinite chain  
model, we need to ensure that the determinant of the infinite dimensional  
matrix under consideration converges. This is, in general, not possible.  
However, if one plots the zeroes of successively larger determinants,  
it is known that in the infinite limit, they converge to the eigenvalues  
of the infinite chain. That is to say that the determinant converges 
on the spectrum of the theory but diverges for $\lambda$ lying between two 
energy levels. For getting the $\beta_n$s one has to repeat the entire
exercise above with the parameters $\mu$ and $\rho$ appropriately 
modified.  

We shall now, as promised, sketch a method for determining the eigenvalues 
of the above model. Towards such an end, let us recast Eq. (34) as 
$$\pmatrix {A_{n+1}\cr A_n} = {\bf R}_n \pmatrix {A_n\cr A_{n-1}}\eqno(51a)$$
where ${\bf R}_n $ is a block upper triangular matrix defined as follows:
$${\bf R}_n = \pmatrix{-{\bf M}_n(E)& -{\bf 1}\cr{\bf 1}& {\bf 0}}\eqno(51b)$$ 
It follows by iterating the map in the above equation that 
$$\pmatrix {A_{n+1}\cr A_n} = {\bf T}_n \pmatrix {A_1\cr A_0}\eqno(52a)$$ 
where
$$ {\bf T}_n =  {\bf R}_n{\bf R}_{n-1}\cdots {\bf R}_1 \eqno(52b)$$  
This is as it should be. The above equation for $A_n$ merely states that 
we can work out all the coefficients if two of them are chosen and if the 
$4\times 4$ transfer matrix ${\bf T}_n $ is known. The matrix ${\bf T}_n$, 
like the matrices ${\bf R}_n, {\bf R}_{n-1}, \cdots {\bf R}_1$, has elements 
which are functions of the variable $E$, which could, at random, take any  
value in the spectrum of the theory. It is well-known that this random matrix 
satisfies the 
Oseledec condition [14] given by 
$$ \lim_{n\to \infty } ({\bf T}_n^\dagger{\bf T}_n )^{1\over 2n} = 
{\bf S}^{-1}{\bf D}{\bf S}\eqno(53a) $$
where 
$${\bf D} = diag~ (\exp {-\Lambda_1} , \exp {-\Lambda_2}, \exp {-\Lambda_3}, 
\exp {-\Lambda_4} )\eqno(53b) $$ 
The $\Lambda_i$ are called the Lyapunov exponents of the infinite chain model 
under consideration and are functions of the random variable $E$.  
If we choose periodic boundary conditions, 
the matrix on the left hand side of Eq.(53a) is one of the many $2nth$ roots  
of the identity matrix {\it i.e.}, it is described by a set of unitary matrices.
As $n\rightarrow \infty$, this set of matrices densely fills up a limiting unit 
circle. It then easily follows that ${\bf S}$ is also a unitary matrix and  
$$\lbrack\lim_{n\to \infty }({\bf T}_n^\dagger{\bf T}_n )^{1\over 2n}
\rbrack^\dagger = {\bf S}^{-1}{\bf D}^*{\bf S}\eqno(53c)$$
Taking the product of the matrices on the left hand 
sides of Eq. (53a) and (53c) and equating it to the product of the 
matrices on the right hand sides we get, 
 $${\bf DD}^* = {\bf 1}\eqno(54)$$
It follows, therefore, that the elements of the diagonal matrix ${\bf D}$
are pure phases. This information can now be fed back into Eqn. (53a).  
The determinant of ${\bf D}$ is purely imaginary and is given by the 
determinant of a unitary matrix on the left hand side of Eqn. (53a). 
The latter, however, is 
not unique. If we pick the simplest root of identity {\it i.e.}, the identity
matrix itself, on the densely filled limiting circle introduced above,  
the matching of the determinants yields 
$$ \Im (\Lambda_1 + \Lambda_2 + \Lambda_3 + \Lambda_4 ) = 2\pi r\eqno(55)$$ 
where $r$ is an arbitrary integer. The above equation restricts the 
randomness in $E$ to some specific functional form in terms of $r$ which  
may be identified with the spectrum of the theory. For other choices of  
the root of identity, the right hand side of the above equation gets shifted
by a real constant. It is difficult to make progress beyond this analytically  
without knowing the Lyapunov exponents.  

We will conclude this paper by summarising the main results and 
outlining some prospective directions for further research. Amongst the 
principal results of this paper is the fact that the infra-red, or the  
zero-momentum, limit of massive non-abelian Chern-Simons theory is described by 
an infinite class of models whose elements are labelled by the angular 
momentum eigenvalue. The first non-trivial model in this class was examined 
in considerable detail and closed expressions for some of the eigenstates 
were obtained in terms of some generalised orthogonal polynomials. 
A method which, in principle, allows us to find the spectrum of the theory 
was also sketched. 
Physically this model corresponds to the coupling of a non-abelian 
Chern-Simons particle with a quantum mechanical non-linear sigma model 
or, equivalently, to a particle moving on the three-sphere - the SU(2) 
group manifold. It would be interesting to see what the models in the
infinite class presented by us correspond to physically for higher angular 
momentum eigenvalues. The two singular points on the three-sphere 
corresponding to $\beta = 0$ and $\beta = \pi$, that we have ignored 
in our analysis, merit discussion in their own right. One has to probably 
use a more refined mathematical method, where one chooses more than one 
coordinate chart to cover the manifold, in order to study this problem. 
It is also probable that following the lines of [6] one may be able 
to derive a phase-space path integral localisation formula for the above 
model. What is perhaps most fascinating is the connection that our 
research opens up between Chern-Simons gauge theories and Anderson's 
chain models.  These models, which are of perennial interest in the 
study of alloys, quasi-crystals, and other disordered systems, are  
a subject of intense investigations in condensed matter physics [12, 13]. 
The assumption we made in Eq. (50) regarding the existence of a zero eigenvalue 
for the chain model is important enough to deserve further 
scrutiny. There exist methods for testing whether the density of states 
in such models is peaked around zero, but the evidence in this regard 
comes from not-so-illuminating numerical work.  
Finally, one can't help wondering if there is a relationship 
between our model and the so-called quasi-exactly soluble models of 
quantum mechanics [15] that have generated some excitement lately. It may be 
recalled that a quasi-exactly soluble quantum mechanical model is one 
in which the model is exactly soluble for a certain range of a 
continuous parameter on which the Hamiltonian depends. In our case,   
instead of a continuous parameter, the Hamiltonian depends on an infinite, 
but  discrete, set of integers which are the angular momentum eigenvalues.  
Although we have studied a non-trivial model associated with only one value 
in this set,
for higher angular momentum eigenvalues, as well as for groups other than 
$SU(2)$, we still expect a discrete spectrum and square integrable 
wavefunctions  because in each case the Hamiltonian is described by a 
self-adjoint operator on a compact manifold. It is however not clear how  
one obtains the eigenstates in these more complicated cases in a closed form.
Indeed, this might well be impossible. We may 
therefore conclude that our model is yet another example - clearly more  
complicated than others hitherto known - of a quasi-exactly soluble model.  
Indeed, at hindsight, the commonality of the important role 
played by the mathematics of orthogonal polynomials in our research 
and that of the subject of quasi-exactly soluble quantum mechanical models
seems more than a mere accident.
\bigskip 
\noindent {${\bf Acknowledgements}$}: This work was moulded into its present
shape by the incorporation of some extremely constructive criticism from 
A. Chatterjee and A. Dasgupta and  
I wish to thank them for the same. I am also happy to acknowledge useful
discussions I had during the course of this work with P. Mitra, A. Niemi,  
L. O'Raifeartaigh, A. Polychronakos, and L. C. R. Wijewardhana.  
\vfil\eject 
\centerline {${\bf References}$}
\item {1. } J. SCHONFELD, {\it Nucl. Phys. } {\bf B185 } (1981), 157;  
S.DESER, R. JACKIW, AND S. TEMPLETON, {\it Phys. Rev. Lett.} 
{\bf 48} (1982), 975; {\it Ann. Phys. (NY)} {\bf 140} (1982), 372; 
C. HAGEN, {\it Ann. Phys. (NY) }{\bf 157} (1984), 342. 
\item {2. } E. WITTEN, {\it Commun. Math. Phys. } {\bf 121} (1989), 351. 
\item {3. } R. PRANGE AND S. GIRVIN, {\it "The Quantum Hall Effect"}, 
Springer-Verlag, NY (1990). 
\item {4. } J. M. LEINAAS AND J. MYRHEIM, {\it Nuovo Cimento} {\bf B37} 
(1977), 1; F. WILCZEK Ed., {\it "Fractional Statistics and Anyon  
Superconductivity"}, World Scientific, Singapore, 1990; G. W. SEMENOFF, 
{\it Phys. Rev. Lett.} {\bf 61} (1988) 517; G. W. SEMENOFF AND P. SODANO, 
{\it Nucl. Phys. }{\bf B328} (1989), 753; A. CHATTERJEE AND V. V. SREEDHAR, 
{\it Mod. Phys. Lett. }{\bf A6} No. 5 (1991), 391; A. CHATTERJEE AND 
V. V. SREEDHAR, {\it Phys. Lett.} {\bf B279} (1992), 69; A. CHATTERJEE,  
R. BANERJEE, AND V. V. SREEDHAR, {\it Ann. Phys. (NY)} {\bf 222} (1993), 254;
M. BOS AND V. P. NAIR, {\it Phys. Lett. }{\bf B223} (1989), 61; S. ELITZUR, 
G. MOORE, A. SCHWIMMER, AND N. SEIBERG, {\it Nucl. Phys. }{\bf B326} (1989), 
108; J. M. F. LABASTIDA AND A. V. RAMALLO, {\it Phys. Lett. }{\bf B227} 
(1989), 92; J. FR\"OHLICH AND C. KING, {\it Commun. Math. Phys. }{\bf 126}
(1989), 167; A. POLYCHRONAKOS, {\it Ann. Phys.} {\bf 203} (1990), 581; 
E. GUADAGNINI, M. MARTELLINI, AND M. MINTCHEV, {\it Phys. Lett.} {\bf B235} 
(1990), 275; {\it Nucl. Phys. }{\bf B336} (1990), 581. 
\item {5. } G. DUNNE, R. JACKIW, AND C. TRUGENBERGER, {\it Phys. Rev. } 
{\bf D41} (1990), 661. 
\item {6. } A. J. NIEMI AND V. V. SREEDHAR, {\it Phys. Lett. } {\bf B336}  
(1994), 381.  
\item {7. } S. DESER AND R. JACKIW, {\it Phys. Lett. }{\bf B139 } (1984), 371; 
P. K. TOWNSEND, K. PILCH, AND P. VAN NIEUWENHUIZEN, {\it Phys. Lett. }
{\bf B136} (1984), 38.
\item {8. } L. D. LANDAU AND E. M. LIFSHITZ, {\it Quantum Mechanics:
Non-Relativistic Theory }  Pergamon Press (1977).  
\item {9. } P. A. M. DIRAC, "Lectures on Quantum Mechanics", Yeshiva 
University Press, New York (1964); L. D. FADDEEV AND R. JACKIW,
{\it Phys. Rev. Lett.} {\bf 60} (1988), 1692.  
\item {10. }  E. GOURSAT {\it A Course in Mathematical Analysis} Vol. {\bf 2}  
Part {\bf  2} translated by E. R. HEDRICK AND O. DUNKEL, Dover Publications,
New York (1959). 
\item {11. } N. I. AKHIEZER, {\it "The Classical Moment Problem" }, 
Oliver \& Boyd, London, (1965).   
\item {12. } P. W. ANDERSON, {\it Phys. Rev. } {\bf 181} (1969), 25. 
\item {13. } R. HAYDOCK, {\it Solid State Physics: Advances in Research
and Applications}, Vol. {\bf 35} ed. H. EHRENREICH, F. SEITZ AND 
D. TURNBULL, Academic Press, New York (1980).  
\item {14. } V. J. OSELEDEC {\it Trans. Moscow Math Soc.} {\bf 19} (1968), 197. 
\item {15. } A. TURBINER, {\it Sov. Phys., J. E. T. P} {\bf 67} (1988), 230; 
A. TURBINER, {\it Comm. Math. Phys.} {\bf 118} (1988), 467; A. MOROZOV, A. 
PERELEMOV, A. ROSLYI, M. SHIFMAN AND A. TURBINER, {\it Int. J. Mod. Phys.}
{\bf A5} (1990), 803; CARL M. BENDER AND GERALD V. DUNNE, {\it "Quasi-Exactly 
Solvable Systems and Orthogonal Polynomials"}, preprint HEP-95-12.  
\vfil\eject\end